\documentclass[osajnl,longbibliography,twocolumn,showpacs,superscriptaddress,10pt]{revtex4-1} 
\usepackage{amsmath,amssymb,graphicx}
\usepackage[final]{hyperref}
\hypersetup{
	colorlinks=true,       
	linkcolor=blue,          
	citecolor=blue,        
	filecolor=magenta,      
	urlcolor=blue         
}

\begin{document}

\title{Propagation of a squeezed optical field in a medium with superluminal group velocity}


\author{Gleb Romanov}
\noaffiliation
\author{Travis Horrom}
\noaffiliation
\author{Irina Novikova}
\noaffiliation
\author{Eugeniy E. Mikhailov}\email[Corresponding author: ]{eemikh@wm.edu}
\affiliation{The College of William $\&$ Mary, Williamsburg VA 23187 USA}

\begin{abstract}

We investigated the propagation of a squeezed optical field, generated via the polarization
self-rotation (PSR) effect, with a sinusoidally-modulated degree of 
squeezing through an atomic medium with anomalous dispersion. We observed the advancement of the signal
propagating through a resonant Rb vapor compared to the reference signal, propagating in air. The measured advancement time grew linearly with atomic
density, reaching a maximum of $11 \pm 1~\mu$s, which corresponded to a
negative group velocity of $v_g\approx - 7,000~$m/s. We also confirmed that the increasing advancement was accompanied by a reduction of output squeezing levels due to optical losses, in good agreement with theoretical predictions. 

\end{abstract}


\maketitle



Manipulations of the group velocity $v_g$ of light using coherent interactions with resonant atoms and atom-like structures have received much attention due to their numerous applications in quantum information, radar steering, all-optical delay lines, etc~\cite{BoydGauthierScience09,milonni_book}.
Numerous experiments have demonstrated that in a ``slow light'' medium (with group index $n_g=c/v_g>1$) both coherent optical pulses and non-classical optical fields are similarly delayed. In particular, single-photon waveforms ~\cite{eisaman_2005} and pulses of squeezed vacuum~\cite{akamatsu_ultraslow_2007,lvovskyPRL08} have been delayed via interactions with Rb atoms in EIT conditions.
However, the propagation of a quantum optical field in a ``fast light'' medium ($n_g<1$) raises some interesting fundamental questions, such as the speed of the information transfer via a superluminal quantum field ~\cite{AharonovPhysRevLett.81.2190,ChiaoPhysRevLett.86.3925,BoydFastlightJO10}. 
Theoretical analysis has predicted that increasing signal advancements must
be accompanied by an unavoidable decrease in the signal to noise ratio thus prevents superluminal information transfer. 
 \begin{figure}[h]
\begin{center}
\includegraphics[width=0.7\columnwidth]{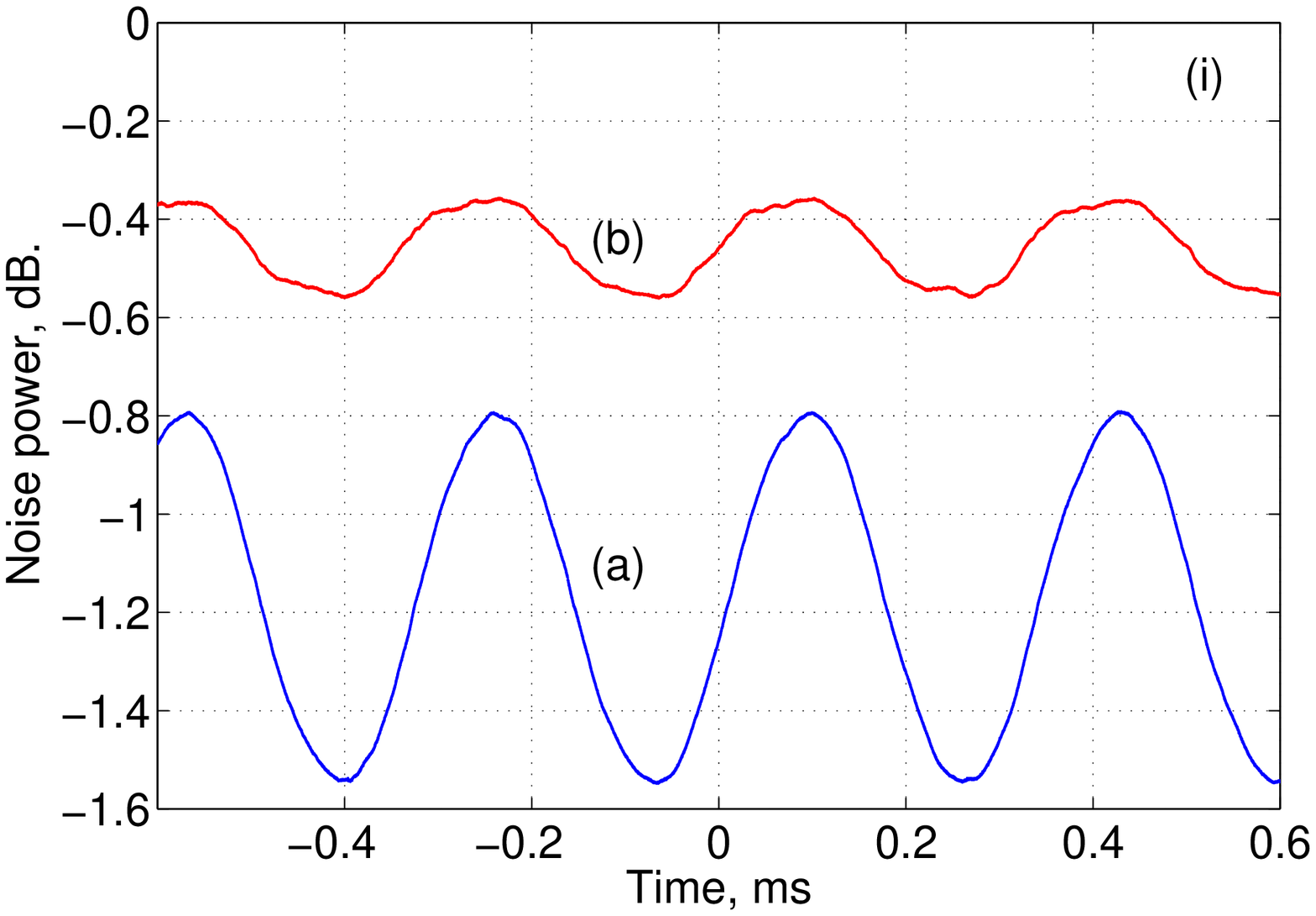}
\includegraphics[width=0.7\columnwidth]{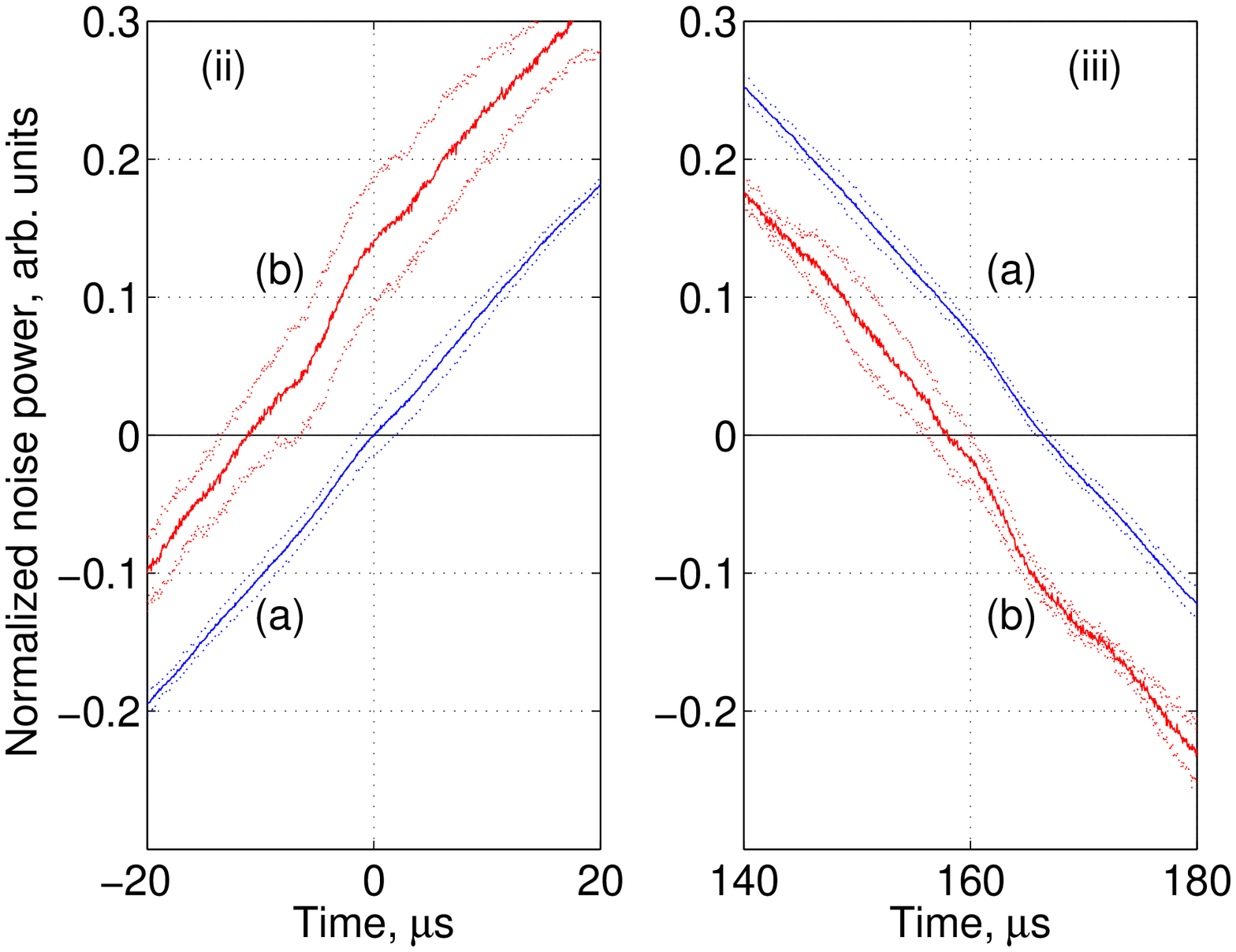}
\caption{
	\label{./mikh_F1} 
	\emph{(i)} Example of the modulated squeezed vacuum noise power of
	the bypass (a) and after interaction with Rb atoms (b). Zero 
	corresponds to the averaged shot noise level. \emph{(ii)} and
	\emph{(iii)} show the zoom-ins of the averaged and normalized
	squeezing traces around the modulation zero crossing.
}
\end{center}
\end{figure} 

Here we study the propagation of squeezed optical field through a ${}^{87}$Rb
vapor cell under fast light conditions due to a nonlinear magneto-optical
interaction~\cite{budker2002rmp}. In these experiments, we let the
sinusoidally-modulated minimum noise quadrature of a squeezed optical field
interact with the Rb vapor, and then compare it with the identically
modulated quantum field propagating in free space. An example of the measurement is shown in Fig.~\ref{./mikh_F1}. Our measurements clearly
demonstrate that the advancement of the quantum noise modulation is due to the interaction with the nonlinear medium. We observed an increasing time shift for both the front and back of the modulation envelope with increased atomic density, accompanied by higher incurred losses for the vacuum field in Rb vapor.

\begin{figure}[htb]
   \begin{center}
   \begin{tabular}{cc}
   \includegraphics[width=0.35\columnwidth]{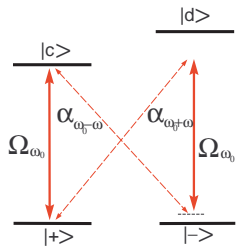}
   \end{tabular}
   \end{center}
   \caption[./mikh_F4] 
   { \label{levels} 
Simplified interaction scheme for PSR squeezing generation. Two
states $|+\rangle$ and $|-\rangle$ represent the orthogonal superpositions
of the ground-state Zeeman substates that are involved in the
interactions of a linearly polarized pump optical field $\Omega_{\omega_0}$
with the hyperfine excited states $|c\rangle$ and $|d\rangle$. 
Here $\omega_0$ is the optical frequency of the pump field.
Two optical fields that close the four-wave mixing loop,
$\alpha_{\omega_0\pm\omega}$,
represent the quantum noise fluctuations of the orthogonally polarized vacuum 
field at the detection frequency $\omega$.
}
   \end{figure}

The modulated squeezed vacuum is produced in the first Rb vapor cell (squeezing cell) via the
polarization self-rotation effect~\cite{matsko_vacuum_2002,ries_experimental_2003}.
This method can be qualitatively described using a simplified four-wave mixing process
~\cite{mikhailov2009jmo} shown in Fig.~\ref{levels}. Due to the difference in the transition
matrix elements, the strong linearly polarized pump field $\Omega_{\omega_0}$ couples the two
hyperfine excited states ($|c\rangle$ and $|d\rangle$) with two orthogonal quantum
superpositions of the ground state Zeeman sublevels ($|+\rangle$ and $|-\rangle$).
The four-wave mixing process, enhanced by the long-lived ground-state Zeeman coherence,
induces correlations between the originally independent quantum fluctuations of the
orthogonally-polarized vacuum field $\alpha_{\omega_0\pm\omega}$, resulting in quadrature
squeezing at the detection frequency $\omega$.
Previous experiments have demonstrated the generation of $\le 3$~dB of broadband low-frequency
squeezed vacuum at several Rb optical resonance frequencies, using only a few mW of pump
laser power~\cite{mikhailov2008ol,grangier2010oe,lezama2011pra}.

The degree of squeezing can be reduced by applying a longitudinal magnetic
field across the Rb cell, without significantly changing the orthogonal (anti-squeezed) quadrature and without rotating the noise ellipse~\cite{HorromJPB12}. This can be qualitatively explained using the
four-wave mixing picture: the presence of the non-zero magnetic field couples the two
ground-states $|+\rangle$ and $|-\rangle$. This deteriorates (or destroys) their mutual
coherence and thus eliminates the correlations between the noise sidebands responsible for squeezing.


   \begin{figure}[htb]
   \begin{center}
   \begin{tabular}{cc}
   \includegraphics[width=0.8\columnwidth]{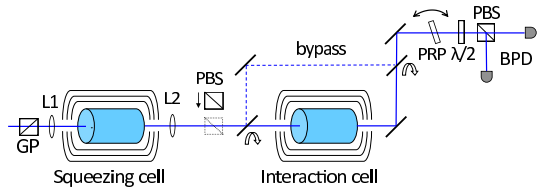}
   \end{tabular}
   \end{center}
   \caption[./mikh_F4] 
   { \label{./mikh_F4} 
Experimental setup. See text for abbreviations. 
}
   \end{figure} 

The general layout for the experiment is shown in Fig.~\ref{./mikh_F4}.
We used an external cavity diode laser tuned and locked  to the ${}^{87}$Rb  $5S_{1/2}F=2\rightarrow 5P_{1/2}F^\prime=1$ transition frequency using saturation absorption spectroscopy. 
We ensured a high quality of the spatial mode of the input laser beam by passing it through a single-mode optical fiber; its linear polarization was controlled using a high quality Glan polarizer (GP). 
Input laser power was $10$~mW.
Both ``squeezing'' and ``interaction'' vapor cells used in the experiments were $7.5$~cm cylindrical Pyrex cells of identical geometry, mounted inside 3$-$layer magnetic shielding. The squeezing cell contained only isotopically enriched ${}^{87}$Rb vapor. The interaction cell, in addition to ${}^{87}$Rb, contained a small amount of Ne buffer gas (2.5~Torr). 

The laser beam was focused inside the squeezing cell using a 30~cm lens
(L1) to the minimum beam diameter of $100~\mu$m, and then recollimated
after the cell with the second lens L2 (focal length $40$~cm) to the
diameter $1.9$~mm. The temperature of the squeezing cell was maintained at
$(66\pm 0.1)^\circ$~C.
To measure the noise quadratures of the output optical field, we employed a detection scheme~\cite{lezama2011pra} that used the strong orthogonally-polarized pump field as a local oscillator without separating it from the squeezed vacuum field.
The relative phase of the polarizations was controlled by tilting a phase retardation plate (PRP) - a quarter-wave plate with its ordinary axis aligned along the pump field orientation.
Then, the polarizations of both optical fields were rotated by $45^\circ$ using a half-wave plate ($\lambda$/2), and evenly split for two inputs of the balanced photodetector (BPD) using a polarizing beam-splitter (PBS), with common noise rejection better than $30$~dB.
We then used a spectrum analyzer to measure the noise of the BPD output,
and observed around $1.6$~dB noise suppression below the shot noise in the range of detection frequencies from $200$~kHz to $2$~MHz.
To experimentally determine the shot noise level, we used another polarizing beam-splitter (PBS) immediately after the squeezing cell, aligned to transmit the pump field and to reject the orthogonally-polarized squeezed vacuum, replacing it with a coherent vacuum.

To measure the group delay we followed the approach similar to previous experiments~\cite{budker99,wang2000nature,mikhailov04josab}; however, instead of monitoring the propagation time of a weak coherent optical probe field, we modulated the degree of squeezing by applying a time-varying magnetic field in the squeezing cell, and then compared the relative shift of the sinusoidal variation at $3$~kHz in the quantum noise propagating through the Rb vapor (in the interaction cell) and in free space (bypass), as shown in Fig.~\ref{./mikh_F4}. 
The modulation amplitude (between 0.8 and 1.5~dB below the shot noise) was chosen so that the noise level of the squeezed vacuum field stays below shot noise at all times. 
The rotation of the pump polarization due to nonlinear Faraday effect did not exceed $2.5$~mrad.
To detect the time dependence on the squeezed quadrature noise power, we utilized  the spectrum analyzer as a narrow-band filter around the detection frequency of $500$~kHz (with a resolution bandwidth of $30$~kHz), and then monitored the video output of the spectrum analyzer on a digital oscilloscope. A sample noise measurement is shown in Fig.~\ref{./mikh_F1}, with each trace consisting of $10^6$ averages.   

\begin{figure}[htb]
\begin{center}
\includegraphics[width=0.8\columnwidth]{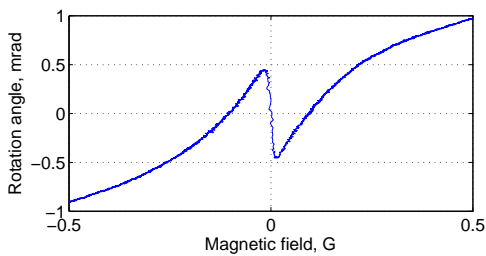}
\caption{
	\label{fig:rot_vs_B} 
	Example polarization rotation signal as a function of longitudinal
	magnetic field $B$ in the interaction cell. Only the central part
	of the wide rotation feature is visible (as an overall positive
	trend), and the region near zero magnetic field is characterized by
	the negative slope due to the narrower rotation feature. The laser
	power is $9.5$~mW at the entrance of the interaction cell.
}
\end{center}
\end{figure}

The collimated output of the squeezing cell, containing both strong pump
field and squeezed vacuum in two orthogonal polarizations, was directed
through the interaction cell. No squeezing occurred in this cell
due to its lower atomic density (temperature) and much lower average laser intensity in 
the unfocussed beam. At the same time, a larger beam size and the presence of the buffer gas increased time-of-flight of atoms through the laser beam. 
We can gain some information about the dispersion properties in the interaction cell by measuring its nonlinear magneto-optical rotation (NMOR) signal ~\cite{budker2002rmp}, since the polarization rotation angle $\phi$ of the linearly polarized optical field is proportional to the magnetically-induced circular birefringence of the atomic medium: 
\begin{equation}\label{eqn_NMOR}
\phi =\frac{L}{2c}\omega_0\left[n_{+}(B)-n_{-}(B)\right]\simeq \frac{L}{c}\omega_0\left(\frac{\partial n_\pm}{\partial \omega}\right)_{B=0} g\mu_B B,
\end{equation}
 where ${\partial n_\pm}/{\partial \omega}|_{B=0}$ is the dispersion for the two circular components for zero magnetic field $B$, $L$ is the length of the atomic medium, $\mu_B$ is Bohr magneton, and $g$ is the gyromagnetic ratio. In the presence of velocity-changing coherence-preserving collisions with a buffer gas, a typical NMOR rotation spectrum, shown in Fig.~\ref{fig:rot_vs_B}, clearly indicates   
two interaction time scales.
The broader rotation slope (characteristic width of approximately 2~MHz) is due to the transient time-limited interaction of light with one optical transition, connected to the slow light propagation~\cite{budker99}.
The narrower feature is due to the atomic diffusion and velocity-changing collisions, resulting in the repeated coherent interactions of light
with the atoms via the both Doppler-broadened excited state hyperfine
components, $F^\prime=1$ and $F^\prime=2$~\cite{NovikovaJOSAB05}. 
Such interaction gives rise to the polarization rotation in the opposite direction (compared to the transient effect), indicating an anomalous dispersion with expected superluminal signal propagation. 
Indeed, the negative dispersion $\frac{\partial n_\alpha}{\partial \omega}<0$ results in
the advancement time $\Delta t_a$ for a weak probe field propagation:
\begin{equation}\label{eqn_advance}
\Delta t_a = \frac{L}{c}-\frac{L}{v_g} \approx  \frac{L}{c}|\omega_0\frac{\partial n_\alpha}{\partial \omega}|.
\end{equation}
To accurately calculate the dispersion for the broadband quantum noise, one has to consider an interaction system, similar to that in Fig.~\ref{levels}, which is beyond the scope of this work. 
Yet, we can use the NMOR spectrum to qualitatively explain the observed advancement for the modulated quantum noise propagation.

%
The typical averaged quantum noise signals before and after interaction with Rb vapor are shown in Fig.~\ref{./mikh_F1}.  
While the quantum noise modulation after the
interaction cell is degraded due to inevitable optical loss, it always stays squeezed, and its shape is well preserved.
The data shown corresponds to the maximum measured advancement of $\Delta t_a = 11 \pm 1~\mu$s for the interaction cell temperature of $(50.0\pm 0.1)^\circ$~C (corresponding to atomic density $1.05\times 10^{11}~\mathrm{cm}^{-3}$~\cite{Rb87numbers}). However, it is hard to directly observe the time difference between the two traces, due to the small value of the fractional delay (limited by the slow modulation period of $>300~\mu$s), as well as due to the difference in the squeezing level due to absorption. To demonstrate the relative advancement more clearly, Fig.~\ref{./mikh_F1}(\emph{ii} and \emph{iii}) shows the normalized modulation signals. The solid curves are the averages of four independent measurements,  and the dotted curves represent two standard deviation boundaries. It is easy to see that the advancement is present both on the leading and trailing fronts of one modulation period for the light traveling through the interaction cell as compared with the bypass. We also observed that the detected time difference was not very sensitive to small variations of the local oscillator phase.


\begin{figure}[htb]
\begin{center}
\includegraphics[width=0.75\columnwidth]{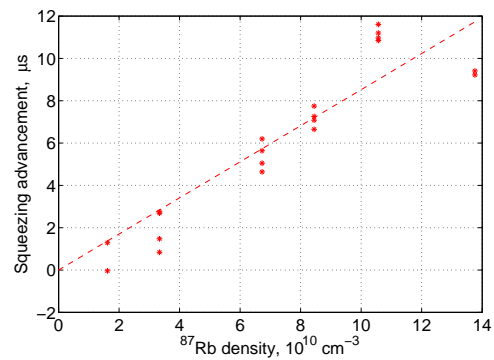}
\caption { 
\label{./mikh_F6} 
	Measured advancement $\Delta t_a$ of the modulated squeezed vacuum
	as a function of the atomic density. The pump laser power before
	squeezing cell is $10$~mW. Each point represents a time difference extracted from fitting the input and output signals with the sine function. The uncertainties of the individual fits are too small to see. 
}
\end{center}
\end{figure} 

To verify that the observed advancement of the modulated quantum noise was due to the interaction with atoms, we repeated the measurements varying the temperature of the interaction cell. For each temperature, we collected several traces to average over day-to-day environmental drifts. Fig.~\ref{./mikh_F6} clearly shows that the observed pulse advancement increases roughly linearly with atomic density (the dashed line represents a linear fit). However, as expected, the squeezing transmitted through the interaction Rb cell deteriorated at higher cell temperature due to increased optical losses.

\begin{figure}[htb]
\begin{center}
\includegraphics[width=0.75\columnwidth]{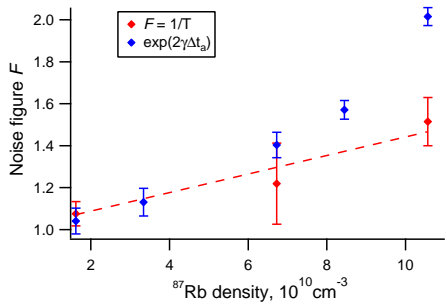}
\caption { 
\label{./mikh_F7} 
	The noise figure $F$, based on estimated transmission of the squeezed vacuum through the interaction Rb cell, and  estimated using Eq.(\ref{fvalue}) from the measured average group delay, shown in Fig.~\ref{./mikh_F6}. 
}
\end{center}
\end{figure} 

Boyd et al.~\cite{BoydFastlightJO10} calculated a simple relationship between the pulse advance $\Delta t_a$ and the noise figure $F$, that is defined as a change in the signal-to-noise ratios before and after the interaction cell, for a simple absorptive resonance of the width $\gamma$ as: 
\begin{equation} \label{fvalue}
F=e^{2 \gamma \Delta t_a}
\end{equation} 
Fig.~\ref{./mikh_F7} shows this predicted average noise figure as a function of atomic density given our measured advancements ($\Delta t_a$) and the approximate  resonance width of $\gamma \approx 2\pi \times 5$~kHz (from Fig.~\ref{fig:rot_vs_B}).  We compare this to the measured noise figure $F=1/T$, as defined in Ref.~\cite{BoydFastlightJO10}. Here $T$ is the transmission coefficient for the squeezed vacuum through the interaction cell, estimated from the experimental data using a beam-splitter model, namely:
\begin{equation} \label{BSPmodel}
\hat{a}_{out}=\sqrt{T}\hat{a}_{in}+ \sqrt{1-T^2}\hat{u},
\end{equation} 
 where the operators $\hat{a}_{in}$ and $\hat{a}_{out}$ represent the input and output optical signal fields, and $\hat{u}$ corresponds to the coherent vacuum mode.
This equation fits the experimentally measured squeezed noise quadratures for input and output reasonably well.
Two points are excluded from Fig.~\ref{./mikh_F7} due to some uncertainty in the local oscillator phase, which affected the detected levels of squeezing (but not the detected time difference).
Even with the limited number of experimental points, it is clear that the increasing advancement in the modulated quantum noise is followed by an increasing noise figure, and a simple model in ~\cite{BoydFastlightJO10} is in reasonably good qualitative agreement with the experimental data.


In conclusion, we have successfully demonstrated the transmission of the
squeezed vacuum through a resonant Rb vapor, in which negative dispersion
was produced via inducing a long-lived Zeeman coherence. We observed that
the modulated quantum noise exits the cell earlier than the analogous
signal traveling in free space, indicating superluminal propagation. The
amount of the advancement increased linearly with the density of atoms.
The largest measured advancement $(11 \pm1~\mu$s) corresponds to a negative
group velocity of $\approx 7,000$~m/s. The increased advancement was accompanied by the deterioration of squeezing due to optical losses, and the measured increase in the noise figure was in the good qualitative agreement with the theoretical predictions of Ref.~\cite{BoydFastlightJO10} and recent experiments with the bright two-mode squeezed twin beams in a ``fast light'' atomic medium ~\cite{LettPRA2012info_negative_velocity,Lettarxiv2013}.
This research was supported by AFOSR grant FA9550-13-1-0098.

%

\end{document}